# EFEKTIFITAS TEKNOLOGI INFORMASI DALAM PROSES BELAJAR MENGAJAR PADA UNIVERSITAS BUDI LUHUR

**Spits Warnars Harco Leslie Hendric**
*Fakultas Teknologi Informasi, Universitas Budi Luhur*
*E-mail: spits@bl.ac.id*

**ABSTRACT**

In general, however, IT will empower students to have greater control over the learning process, with all the benefits associated with active learning and personal responsibility. Not only will students decide when to learn and how to learn, increasingly they will also decide what to learn and how that learning is to be certified. Traditionally, higher education institutions have combined several functions in their faculty. Faculty are architects as they design learning programs; navigators as they help advise students in their course of study; instructors when they lecture; mentors when they help students form a sense of connectedness to the world; and evaluators and certifiers as they decide to grant students grades or degrees.

## 1. PENDAHULUAN

Dewasa ini teknologi informasi (TI) sudah bisa dikatakan memegang peranan penting dalam aspek kehidupan manusia. Dengan teknologi informasi melalui layanan internet maka informasi tersebut dapat tersebar dengan sangat cepat dan tanpa batas. Dalam dunia usaha, teknologi informasi tersebut sudah menjadi hal umum untuk dipergunakan sebagai penunjang bisnis dalam kegiatan operasional maupun kegiatan yang bersifat strategis. Sekarang ini dalam lingkup pendidikan, TI bukan hanya dipergunakan institusi pendidikan seperti halnya perguruan tinggi sebagai pendukung administratif saja tapi juga dipakai dalam kegiatan proses pembelajaran yang dipandang sebagai hal yang strategis dalam membantu mencapai visi, misi dan tujuan dari institusi tersebut.

Universitas Budi Luhur sebagai salah satu universitas yang dikenal sebagai pendiri awal program studi yang berbau teknologi informasi, dengan mendirikan Akademi Ilmu Komputer (AIK) Budi Luhur pada tanggal 1 April 1979 hingga perkembangannya sampaiga saat ini. Universitas Budi Luhur merasa penting untuk mengedepankan teknologi informasi dalam proses belajar mengajar, walau dikatakan terlambat UBL berusaha mengejar ketinggalan dengan memulai menggunakan teknologi informasi sejak tahun 2003. Walaupun yang dirintis masih sebatas penggunaan alat bantu komputer, LCD untuk mendukung proses belajar mengajar.

Sebelum kita masuk lebih dalam kedalam pembahasan ada baiknya kita definisikan proses belajar mengajar secara tradisional dan proses belajar mengajar dengan teknologi informasi. Dijelaskan proses belajar mengajar yang secara traditional, seperti tatap muka, yang dilakukan pada waktu yang telah ditentukan dan tempat (ruang kelas) yang juga sudah ditentukan. Proses belajar mengajar tradisional yang dimaksud disini juga masih hanya mempergunakan papan tulis dengan kapur atau marker/spidol sebagai alat tulisnya dan makalah serta buku-buku yang terbuat dari kertas sebagai acuan dan panduan dalam proses belajar mengajar. Termasuk juga penggunaan transparasi dengan cara menuliskan bahan kuliah pada plastik transparansi dengan menggunakan spidol khusus, dan ditampilkan di ruang kelas pada overhead projektor.

Proses belajar mengajar dengan Teknologi Informasi yaitu adanya peranan teknologi informasi pada proses belajar mengajar perguruan tinggi. Penggunaan Teknologi untuk mengajar. Disini mungkin penerapan teknologi yang masih sebahagian saja, kegiatan tatap muka masih dilakukan namun penyampaian materi kuliah dibantu oleh teknologi seperti penggunaan power point menggantikan papan tulis, penyampaian materi dengan audio video, atau dengan software interaktif di depan komputer.

Dilihat dari sisi pengembangan teknologi yang mungkin mempunyai dampak terhadap proses belajar mengajar dilakukan dengan mempelajari bagaimana internet itu berkembang, bagaimana software-software yang baru bermunculan dapat berpengaruh pada mahasiswa, interaksi antar mereka baik secara bilateral maupun berdasarkan kolektif, bagaimana mereka memperoleh dan meyimpan informasi

## 2. KEUNTUNGAN PENERAPAN TEKNO-LOGI INFORMASI DALAM PROSES BELAJAR MENGAJAR

Menurut Higgins A. (2001) ada beberapa kegiatan untuk mendukung penerapan TI pada proses belajar mengajar dan akhirnya menimbulkan keuntungan dalam menggabungkan TI dalam proses belajar mengajar, yaitu:
1. Mahasiswa membuat dan mengirim tugas secara elektronik
   Mahasiswa dapat ditugaskan membuat tugas dengan software pengolah kata seperti

Microsoft Word atau spreadsheet seperti Microsoft Excell. Tugas ini dapat dikirm melalui email,disket, beberapa keuntungan tugas elektronik ini yaitu:
   a. Tidak membutuhkan kertas
   b. Waktu dan tanggal pengerjaan tercatat
   c. Format standar
   d. Mudah dalam penyimpanan
   e. Dapat memantau mahasiswa yang sedang aktif
2. Gunakan test online untuk mengelola dan menilai objektif test
   Online test ini mempunyai keuntungan:
   a. Otomatis penghitungan grade
   b. Penelusuran kemajuan mahasiswa
   c. Kemampuan database dan penyimpanan
   d. Pengurangan biaya fotocopy dan kertas
   e. Mahasiswa dapat secara langsung melihat hasilnya
3. Virtual komunikasi
   Menggunakan mailing list atau web conferences untuk berkomunikasi dengan mahasiswa, mahasiswa dengan mahasiswa lainnya, membuat grup diskusi virtual dan promosikan kolaborasi mahasiswa. Keuntungan dari penggunaan virtual komunikasi ini mempunyai jangkauan luas yaitu:
   a. Keterbatasan waktu pertemuan kelas dan jumlah mahasiswa perkelas yang besar dapat teratasi, yang memungkinkan untuk mengelola grup diskusi dengan benar
   b. Memberi kesempatan pada semua mahasiswa
   c. Memaksa mahasiswa untuk menyusun pemikiran mereka dalam sebuah tulisan
   d. Menerima lebih banyak umpan balik berkenaan dengan harapan, tujuan atau kepuasan mahasiswa
   e. Menyediakan umpan balik pada mahasiswa lebih efisien dan mengikutsertakan dosen tamu tanpa harus menghadirkan dosen tamu
   f. Mengkumpulkan grup mahasiswa yang tidak dapat belajar bersama, contoh mengkombinasikan kolega mahasiswa luar negeri dengan mahasiswa yang mempunyai diskusi yang sama.
4. Tampilkan materi, catatan dan instruksi kuliah pada web
   Cara ini mempunyai keuntungan:
   a. Mahasiswa lebih mudah mengakses materi, catatan dan instruksi kuliah
   b. Menyediakan acuan yang jelas untuk sebuah instruksi yang khusus
   c. Menghemat waktu dalam menjelaskan sebuah materi kuliah
   d. Membuat hubungan antara materi kuliah yang terbagi menjadi kursus yang anda ajar.
   e. Menyediakan sumber web yang releven dengan mata kuliah
5. Menggunakan multimedia di ruang kelas
   Dapat dilakukan dengan menampilkan video atau animasi untuk menggambarkan penjelasan mata kuliah, ataupun dengan menggunakan software presentasi seperti Microsoft powerpoint untuk menampilkan catatan kuliah. Dengan multimedia ini kita dapat:
   a. Menggunakan suara,gambar atau animasi untuk mendukung pembelajaran aktif dan membantu mahasiswa mempelajari secara abstrak
   b. Mengorganisasikan catatan kuliah
   c. Menyimpan catatan kuliah secara digital untuk keperluan akan dating
   d. Mengurangi biaya plastik presentasi dan tinta untuk pembuatan presentasi
   e. Menggunakan catatan kuliah untuk kepentingan lain
6. Menggunakan jurnal elektronik dan sumber akademik online lainnya
   Setiap hari materi jurnal dan akademik tersedia pada web, beberapa harus bayar dan beberapa bebas biaya. Penggunaan materi elektronik ini membantu
   a. Mengurangi biaya foto kopi
   b. Mengurangi pengaturan teks secara manual

**3. KEBERHASILAN PENERAPAN TEKNO-LOGI INFORMASI**

Luftman (2004, 9) mengatakan bahwa keberhasilan dan kegagalan TI dapat lebih kelihatan ketika mempengaruhi reputasi organisasi dan operasionalnya meluas. Demikian juga dengan penggunaan TI untuk mendukung proses belajar mengajar akan dikatakan berhasil atau gagal berdasarkan reputasi dan operasional institusi pendidikan yang menerapkannya.

Suatu universitas dikatakan berhasil menerapkan TI jika:
1. Dikenal masyarakat karna menggunakan dan menerapkan teknologi baik dalam proses belajar mengajar maupun administrasinya
2. Dibukanya program studi baru dalam rangka perluasan operasional, dimana pembukaan program studi baru tersebut dibantu oleh data warehouse sebagai alat bantu untuk menentukan keputusan manajemen tingkat atas dalam membuka sebuah program studi baru.
3. Dibukanya cabang universitas di tempat lain dalam rangka perluasan operasional, dimana pembukaan cabang baru tersebut dibantu oleh data warehouse sebagai alat bantu untuk menentukan keputusan manajemen tingkat atas dalam membuka sebuah cabang baru.
4. Memuaskan mahasiswa baik dalam proses administrasi kuliah maupun perkuliahan, dimana TI membantu mahasiswa untuk

a. Lulus kuliah tepat waktu, proses belajar-mengajar yang terbantu oleh TI membantu mahasiswa yang ketinggalan perkuliahan. Seperti yang diutarakan oleh Higgins A. (2001) ada beberapa keuntungan dalam menggabungkan TI dalam proses belajar mengajar.
b. Cepat dan mudah administrasi kuliah
5. Meningkatkan penerimaan mahasiswa baru
6. Mendapatkan akreditasi A, dimana akreditasi ini dikeluarkan oleh KOPERTIS sebagai lembaga wadah perguruan tinggi swasta yang salah satu tugasnya membantu masyarakat untuk menentukan kualitas suatu institusi pendidikan tinggi di Indonesia. Pembuatan laporan akreditasi ini dapat dibantu oleh datawarehouse yang berfungsi untuk menyimpan data-data yang bersifat strategis.

## 4. PENGHAMBAT PENERAPAN TEKNOLOGI INFORMASI

Luftman (2004, 34) juga berpendapat bahwa TI dapat menjadi penghambat transformasi bisnis yang dapat dilakukan melalui beberapa mekanisme berikut:
1. Ketika strategi TI tidak selaras dengan strategi bisnis
2. Penekanan teknologi yang berlebihan oleh manajemen TI dan manajemen bisnis
3. Kegagalan manajemen TI dan bisnis untuk mengenali efektifitas penggunaan TI untuk merubah proses bisnis.

Berdasarkan konsep Luftman diatas penting bagi perguruan tinggi untuk menghindarkan hal-hal yang disampaikan oleh Luftman berkenaan dengan penghambat transformasi bisnis, sehingga penerapan TI tidak menjadi penghambat.

Selain itu dalam proses belajar mengajar para pengajar (dosen) harus memiliki motivasi yang sangat besar apabila teknologi akan diterapkan di ruang kelas (Gahala, 2001). Banyak hambatan-hambatan yang akan bermunculan. Teknologi bisa saja sangat mengintimidasi bagi sebahagian para dosen karena untuk memperkenalkan teknologi kepada mereka juga membutuhkan proses pembelajaran tersendiri. Para pengajar ini mungkin kekurangan waktu dan motivasi untuk mempelajari keahlian teknologi informasi.

## 5. KESIMPULAN

Berdasarkan uraian diatas dapatlah ditarik kesimpulan:
1. Penggunaan TI pada proses belajar pada Universitas Budi Luhur akan berhasil jika reputasi organisasi dan operasionalnya meluas.
2. Penggunaan TI pada proses belajar pada Universitas Budi Luhur akan gagal jika reputasi organisasi dan operasionalnya tidak meluas
3. Penerapan TI pada Universitas Budi Luhur akan terhambat jika strategi TI tidak selaras dengan strategi bisnis, penekanan teknologi yang berlebihan oleh manajemen TI dan bisnis, serta manajemen TI dan bisnis gagal mengenli efektifitas penggunaan TI untuk merubah proses bisnis.
4. Universitas Budi Luhur akan maju sebagai Universitas yang maju dan bersaing jika menerapkan beberapa kegiatan yang disarankan oleh Higgins A.(2001), yang pada akhirnya akan memberikan keuntungan pada Universitas Budi Luhur terutama dalam proses belajar mengajar.